\documentclass[aps,prl,reprint,groupedaddress,showpacs]{revtex4-2}
\usepackage{amsmath}
\usepackage{amssymb}
\usepackage{graphicx}
\usepackage{array}
\usepackage{dcolumn}
\usepackage{subfigure}
\usepackage{color}
\usepackage{float}
\usepackage{xr-hyper}
\usepackage[hidelinks=true]{hyperref}
\hypersetup{
  colorlinks   = true, 
  urlcolor     = blue, 
  linkcolor    = blue, 
  citecolor   = red 
}

\begin{document}

\title{Latent electronic (anti-)ferroelectricity in BiNiO$_3$}
\author{Subhadeep Bandyopadhyay}
\email{subha.7491@gmail.com}
\author{Philippe Ghosez}
\email{Philippe.Ghosez@uliege.be}
\affiliation{Theoretical Materials Physics, Q-MAT, University of Liège, B-4000 Sart-Tilman, Belgium}

\begin{abstract}
BiNiO$_3$  exhibits an unusual metal-insulator transition from $Pnma$ to $P\overline{1}$ that is related to charge ordering at the Bi sites, which is intriguingly distinct from the charge ordering at Ni sites usually observed in related rare-earth nickelates. Here, using first principles calculations, we first rationalize the phase transition  from $Pnma$ to $P\overline{1}$, revealing an overlooked intermediate $P2_1/m$ phase and a very unusual phase transition mechanism. Going further, we point out that the  charge ordering at Bi sites in the $P\overline{1}$ phase is not unique. We highlight an alternative polar orderings giving rise to a ferroelectric $Pmn2_1$ phase nearly degenerated in energy with $P\overline{1}$ and showing an in-plane electric polarisation of 53 $\mu C $/cm$^2$ directly resulting from the charge ordering.  The close energy of $Pmn2_1$ and $P\overline{1}$ phases, together with low energy barrier between them, make BiNiO$_3$ a potential electronic antiferroelectric in which the field-induced transition from non-polar to polar would relate to non-adiabatic inter-site electron transfer. We also demonstrate the possibility to stabilize an electronic ferroelectric ground state from strain engineering in thin films, using an appropriate substrate.

\end{abstract}

\date{\today}

\maketitle

Nickelate perovskites ($R$NiO$_3$ with $R$ = Y or a rare-earth element) have generated a significant interest over the last years due to their fascinating electronic, magnetic and structural properties, potentially linked to a wide variety of functional applications\cite{appl_nick_1,appl_nick_2}. $R$NiO$_3$ compounds (except $R$=La) undergo a metal to insulator transition (MIT)  with associated  structural phase transition from  high-temperature orthorhombic $Pnma$ to low-temperature monoclinic $P2_1/n$ structure \cite{PT_RNO1}.  The critical temperature of the MIT decreases with increasing $R^{3+}$ ionic radius and is finally suppressed for LaNiO$_3$, which exhibits a distinct metallic $R\overline{3}c$ phase at all temperatures. For smaller $R^{3+}$ ion, the MIT is driven by a breathing distortion of the NiO$_6$ octahedra, which creates two inequivalent Ni sites and subsequent charge ordering (CO), $ 2 \rm{Ni}^{3+} \rightarrow \rm{Ni}^{2+}+\rm{Ni}^{4+}$ \cite{Ni_disp,PT_RNO2}. At the electronic level, considering Ni-O hybridizations, this formal transition is often better reformulated in terms of oxygen holes ($L$):  $2({\rm Ni}^{2+}L^1) \rightarrow {\rm Ni}^{2+}+ {\rm Ni}^{2+}L^2$ \cite{O_hole1,O_hole2}. At the structural level, it has been shown that the breathing distortion is triggered by the oxygen octahedra rotations (OOR) inherent to the $Pnma$ phase \cite{YNO_Mercy}. This behavior is ubiquitous amongst the $R$NiO$_3$ compounds, making them a distinct and well-defined family of materials.

We might naturally expect BiNiO$_3$ to belong to this class of compounds. In view of the similar size of Bi$^{3+}$ and La$^{3+}$ cations, it is questionable why BiNiO$_3$ does not behave like LaNiO$_3$ \cite{Shanon,Radii}. However, relying instead on bond-valence analysis \cite{BVpa}, it appears that BiNiO$_3$ has a Goldschmidt tolerance factor \cite{Goldschmidt} very similar to SmNiO$_3$ (see Fig.S1(a)). In line with that, BiNiO$_3$ shows a metallic $Pnma$ phase with OOR amplitudes comparable to those of SmNiO$_3$ (see SI). Like the latter, it then exhibits an insulating ground state but instead of crystallizing in the same insulating $P2_1/n$ phase with CO at the Ni sites, it is reported in an unusual $P\overline{1}$ phase combining an unexpected Ni$^{2+}$ state with CO at the Bi sites ($\rm{Bi}^{3+} \rm{Ni}^{3+} \rightarrow \rm{Bi}^{3+}_{1/2} \rm{Bi}^{5+}_{1/2}\rm{Ni}^{2+}$) \cite{BNO_02,BNO_JACS}. Although the $P2_1/n$ phase has been theoretically predicted to be metastable \cite{twotypes_BNO}, it has never been experimentally observed. A temperature versus pressure phase diagram has been reported experimentally, suggesting direct phase transition from $Pnma$ to $P\overline{1}$ at a critical temperature decreasing linearly with increasing pressure~\cite{BNO_NAT}.

Various studies have discussed the MIT in BiNiO$_3$, focusing mainly on the electronic properties. Dynamical mean field theory calculations~\cite{BNO_Hartree,BNO_DMFT} reproduce CO of Bi$^{3+}$ and Bi$^{5+}$ in the insulating phase assuming Bi$^{4+}$ to be a valence skipper with an attractive Hubbard interaction, while the formal Bi$^{3+}$Ni$^{3+}$ occupancy makes BiNiO$_3$ a metal in the $Pnma$ phase. This integer valence description is too simple to reflect the exact electronic configurations and X-ray absorption spectroscopy finds a charge state away from  Ni$^{3+}$~\cite{xRAY_BNO} in the metallic state. Paul {\it et al.}~\cite{BNO_PRL} then better proposed an description of the form $({\rm Bi}^{3+} L^{\delta}) ({\rm Ni}^{2+} L^{1-\delta}) \rightarrow {\rm Bi}^{3+}_{1/2} ({\rm Bi}^{3+} L^{2(1-\delta)})_{1/2} ({\rm Ni}^{2+}L^{\delta})$ involving oxygen holes $L$ and explained the pressure dependence of the MIT from changes of Bi-O and Ni-O hybridizations. Although this alternative view is likely more accurate, we continue hereafter using an integer description of Bi valence that provides a simplified but qualitatively correct global picture.

Here, we report a detailed first-principles study of BiNiO$_3$ addressing together electronic and structural aspects. Our approach accurately reproduces the CO and $P\overline{1}$ ground state. First, we unveil the existence of an intermediate $P2_1/m$ phase along the path from the high-temperature  $Pnma$ phase to  the $P\overline{1}$ ground state and an unusual transition mechanism from $P2_1/m$ to $P\overline{1}$ involving only stable modes. Then, we point out that the CO of the $P\overline{1}$ phase is not unique and identify an alternative CO giving rise to a ferroelectric $Pmn2_1$ phase of comparable energy. We clarify that ferroelectricity in that phase is electronic in nature and discuss practical implication of our findings in terms of electronic (anti-)ferroelectricity. 

\begin{figure}[h!]
\includegraphics[width=\columnwidth]{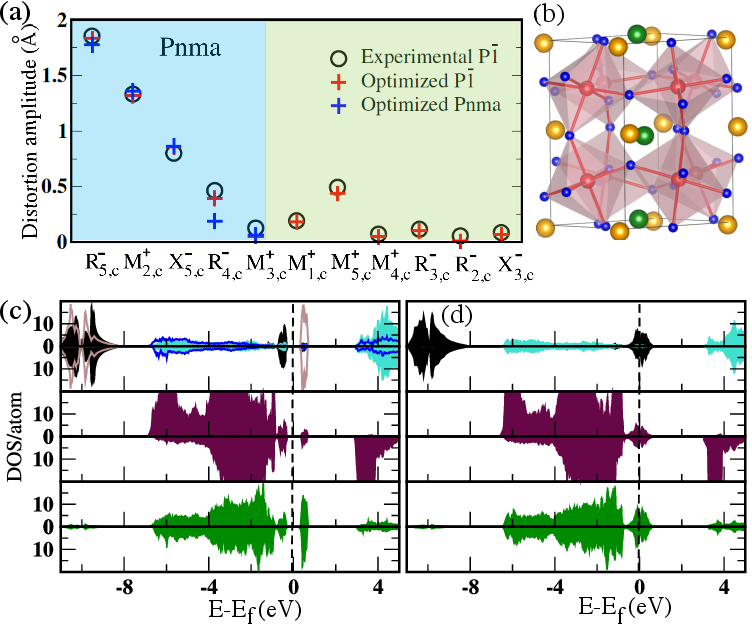}
\caption{(a) Symmetry-adapted mode analysis of the atomic distortion, respect to the cubic phase, in the experimental (open circle) \cite{BNO_attfield} and optimized ({\color{red} +}) $P\overline{1}$ structures and in the optimized ({\color{blue} +}) $Pnma$  phase. (b) Sketch of the $P\overline{1}$ crystal structure, with Bi$_S=$ Bi$^{5+}$ (green), Bi$_L=$ Bi$^{3+}$ (orange), Ni (red) and O (blue) atoms. Partial DOS of (c) $P\overline{1}$ and (d) $Pnma$ phases, highlighting Bi$_L$ $6s$ (black), Bi$_S$ $6s$, (brown line), Bi$_L$ $6p$ (cyan), Bi$_S$ $6p$ (blue line ), Ni $3d$ (maroon), O $2p$ (green) contributions. Vertical dashed  line locates $E_f$.}
\label{P_1}
\end{figure}

Our calculations are performed using a DFT+U approach, relying on the PBEsol~\cite{PBESOL} exchange-correlation functional, as implemented in ABINIT software\cite{ABINIT2,ABINIT3,ABINIT4} (see SI). 
U and J  corrections are included for Ni 3$d$ states~\cite{U}. We checked the results for different (U,J) values and found that (6,1) eV  provides excellent theoretical description of the experimental  $P\overline{1}$ ground state (see SI Table.ST1 and Fig.\ref{P_1}(a)). 
For too small U,  $P\overline{1}$ cannot be stabilized, consistently with Ref.\cite{twotypes_BNO}. Symmetry-adapted mode analysis is performed using Isodistort \cite{Isodistort}. The phase to which refer each symmetry label is identified with a subscript : $c$ for cubic ($Pm\bar{3}$m), $o$ for orthorhombic ($Pnma$) and $m$ for monoclinic ($P2_1/m$). Connection between symmetry labels of the three phases is reported in Table.ST3.
Non adiabatic charge transfer is probed using constrained DFT, as implemented in ABINIT \cite{cDFT}.


$\bf{P\overline{1}}$ $ \bf {ground~ state}$ --
Starting from the experimental $P\overline{1}$  structure, we first carry out full structural optimization for different collinear magnetic configurations of Ni. Comparing ferromagnetic (FM) with A, C and G type antiferromagnetic (AFM) spin orders, we find G-AFM to be energetically the most favorable order (Table.ST2), with a theoretical unit cell volume (233 \AA$^3$) comparable to experiment (233-235 \AA$^3$)~\cite{BNO_JACS,BNO_NAT}. Since G-AFM spin configuration also remains the most favorable in other phases, it is kept all along this work.

Relying on symmetry-adapted mode analysis \cite{Isodistort}, we point out in Fig.\ref{P_1}a the excellent agreement between the  optimized and experimental \cite{BNO_attfield} atomic distortions of the $P\overline{1}$ structure, with respect to the $Pm\overline{3}m$ cubic  reference. Amongst these distortions, some are already inherent to the intermediate $Pnma$ phase \cite{RTiO3_modes,YNO_Mercy}: primary in-phase ($M_{2,c}^{+}$) and anti-phase ($R_{5,c}^{-}$) NiO$_6$ octahedra rotations together with secondary anti-polar motions of Bi atoms ($X_{5,c}^{-}$ and $R_{4,c}^{-}$) and more negligible Jahn-Teller distortion ($M_{3,c}^{+}$). 
Then, additional $M_{1,c}^{+}$ and $M_{5,c}^{+}$ distortions (Fig.S3) are also present, which explain together the lowering of symmetry from $Pnma$ to  $P\overline{1}$ : $M_{1,c}^{+}$ motions of O atoms in ab-plane, which induce a breathing-like distortion of BiO$_{12}$ polyhedra and $M_{5,c}^{+}$ anti-phase motions of O atoms along c, which distort the polyhedra further. This gives rise to large (Bi$_L$, 51.07 \AA$^3$) and small (Bi$_S$, 47.24 \AA$^3$) Bi sites that order according to a C-type pattern in which Bi$_L$ and Bi$_S$ alternate along two directions and are preserved in the third one (Fig.\ref{P_1}b). Small  $R_{3,c}^{-}$, $M_{4,c}^{+}$ and $X_{3,c}^{-}$ distortions are also present. The negligible contribution of the $R_{2,c}^{-}$ mode confirms the absence of breathing distortion at Ni sites,  dominant in the insulating $P2_1/n$ phase of other $R$NiO$_3$ perovskites \cite{PT_RNO2,YNO_Mercy}. 

\begin{figure}[h!]
\includegraphics[width=\columnwidth]{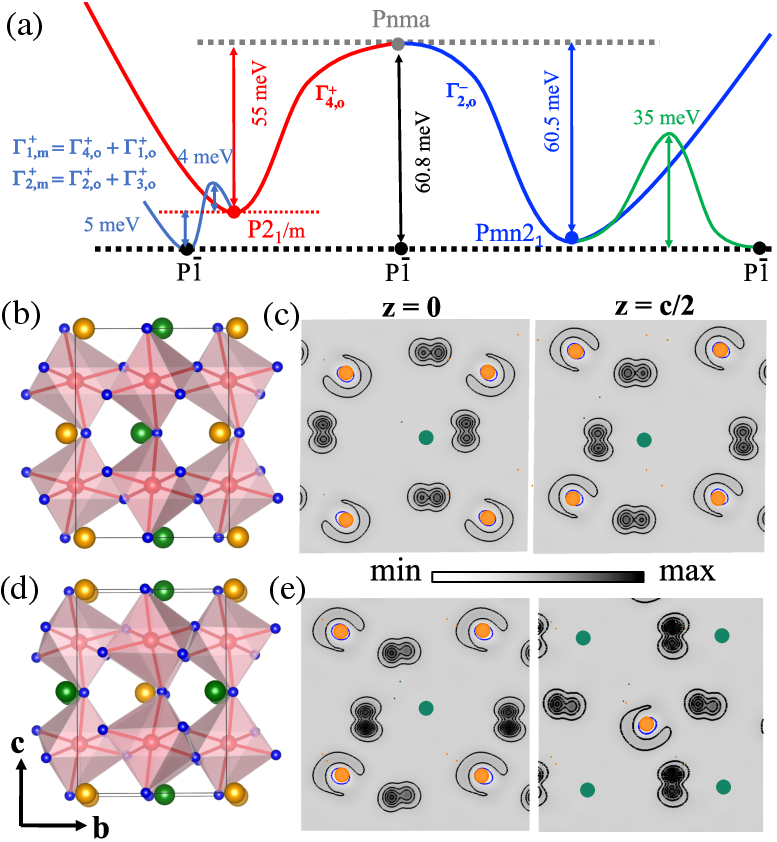}
\caption{(a) Schematic energy landscape of BiNiO$_3$, locating the different phases and the energy barriers between them (G-type AFM spin ordering). Crystal structure and partial charge density of top valence electrons (b, c) in the  $P2_1/m$ and $P\overline{1}$ phases and (d, e) in the $Pmn2_1$ phase.}
\label{metastable}
\end{figure}

The partial density of states (PDOS) in Fig.\ref{P_1}(c) reveal dominant antibonding Bi $6s$ + O $2p$ contributions around the Fermi energy (E$_f$), whereas bonding states are lying much deeper (i.e. $\sim $10 eV below E$_f$). In the $P\overline{1}$ phase, a splitting between antibonding Bi $6s$ + O $2p$ states is opening a band gap of 0.5 eV in line with experiment\cite{Gap_BNO}. Distinct Bi$_L$ and Bi$_S$ contributions with occupied (unoccupied) $6s$ levels near E$_f$ are consistent with Bi$^{3+}$ and Bi$^{5+}$ (or Bi$^{3+}\underline{L}^2$) states, giving rise to CO according to a C-type pattern \cite{BNO_PRL}. This is confirmed by charge density plots of top valence electrons (Fig.S4a, Fig.\ref{metastable}c), highlighting the presence of a Bi $6s$ lone pairs at Bi$_L$ site only. These lone pairs are pointing along the pseudo-cubic diagonal in each $ab$-plane; they are lying on the same side of Bi atoms in a given $ab$-plane and in opposite sides in consecutive layers, in line with anti-polar motion of Bi atoms and inversion symmetry of the system. 
 
Also, PDOS of the Ni-$3d$ show that $t_{2g}$ states are occupied for both the spin channel and $e_g$ states are occupied (empty) for majority (minority) spin channel. This confirms a high-spin Ni$^{2+}$($t_{2g}^6e_g^2$) state, consistent with the calculated magnetic moment of $\sim$1.67 $\mu_B$/Ni. Small differences in the Ni magnetic moments results in an uncompensated ferrimagnetic (FiM) net magnetization of 0.01 $\mu_B$. Such a weak magnetisation is also observed experimentally, but as a result of a canted G-AFM ordering\cite{BNO_attfield}.
\par
$\bf{Pnma}$  $\bf{phase}$ --
 The $Pnma$ phase lies 61 meV/f.u higher in energy than $P\overline{1}$. 
Its relaxed unit cell volume (228 \AA$^3$) is $\sim$2.4 \% smaller than that of $P\overline{1}$, consistently with the $\sim$2.5 \%  volume shrinkage observed experimentally during the $P\overline{1}-Pnma$ transition at 3.5 GPa\cite{BNO_NAT}. 
Structurally, the $Pnma$ phase ($a^-b^+c^-$ in Glazer's notations) shows large out-of-phase and in-phase NiO$_6$ octahedra rotations of 9.6$^{\circ}$ and 11.2$^{\circ}$, which remain similar in the $P\overline{1}$ phase (Fig.\ref{P_1}a). At the electronic level, the PDOS (Fig.\ref{P_1}d) point out a metallic character, with partially occupied Bi $6s$ and O $2p$ antibonding states at $E_f$. The significant occupancy of Ni $3d$ states and the Ni magnetic moment of 1.65 $\mu_B$ indicate a charge state closer to high-spin Ni$^{+2}$ than to Ni$^{+3}$, in line with experimental observations \cite{xRAY_BNO}. Consequently, the nominal charge state of Bi should be Bi$^{4+}$, which suggests a strong tendency to  electronic instability since Bi$^{4+}$ is a valence skipper\cite{skipping}. Accordingly, the $Pnma$ phase shows two unstable phonon modes at $\Gamma$ : a  $\Gamma_{4,o}^{+}$ ( 310$i$ cm$^{-1}$) and a $\Gamma^{-}_{2,o}$ mode (149$i$ cm$^{-1}$) that both induce CO at Bi sites.

Condensing the $\Gamma_{4,o}^+$ mode lowers the symmetry to $P2_1/m$ and give rise to a relaxed insulating metastable phase located 55 meV/f.u. below the $Pnma$ phase (Fig. \ref{metastable}a). Inspection of the PDOS highlights a bandgap of 0.46 eV and confirms charge disproportionation at the Bi sites (Fig.-S5). This $P2_1/m$ phase shows a C-type CO and lone-pair orientations similar to $P\overline{1}$ (Fig. \ref{metastable}b-c).

Condensing instead the $\Gamma^{-}_{2,o}$ mode lowers the symmetry to $Pmn2_1$ and give rise to another insultating metastable phase located 60.5 meV/f.u. below the $Pnma$ phase (i.e. only 0.3 meV/f.u. above $P\overline{1}$).  
Inspection of the PDOS also show charge disproportionation at the Bi sites (shown in Fig.S5) but Bi$_L$ 6s states are much broaden at the conduction level (compared to $P2_1/m$ and $P\overline{1}$), indicating stronger Bi $6s$ -- O $2p$ hybridizations resulting in a smaller bandgap of 0.3 eV. Moreover, Bi$^{3+}$ and Bi$^{5+}$ sites now alternate along the three directions giving rise to a G-type CO that breaks the inversion symmetry in line with the polar character of the $Pmn2_1$ phase. 

Interestingly, the only appearance of C-type (resp. G-type) CO in $Pnma$ already lowers the symmetry to $P2_1/m$ (resp. $Pmn2_1$). Together with the close energies of $Pmn2_1$,  $P2_1/m$ and $P\overline{1}$ phases, this emphasizes that the major driving force destabilizing the $Pnma$ structure is CO, whatever the resulting order.

{\bf Path to the ground state} -- Amazingly the $P2_1/m$ and $Pmn2_1$ phases are both dynamically stable. The natural path from $Pnma$ to $P\overline 1$ should preferably go through $P2_1/m$, which already condense $\Gamma_{4,o}^+$ distortion. In the monoclinic $P2_1/m$ phase, none of the mode is however unstable but additional condensation of the low frequency $\Gamma^{+}_{2,m}$ mode (50 cm$^{-1}$) properly brings the system to the $P\overline 1$ ground state. Doing so requires however to overcome an energy barrier of 4 meV/f.u.. 

In order to clarify the mechanism of this unusual phase transition condensing a stable mode, we studied the energy landscape around the $P2_1/m$ phase from a Landau-type expansion (up to 4$^{th}$ order) involving $\Gamma^{+}_{2,m}$=$\Gamma^{+}_{3,o}$ $\oplus$ $\Gamma^{+}_{2,o}$ and $\Gamma^{+}_{1,m}$=$\Gamma^{+}_{1,o}$ $\oplus$ $\Gamma^{+}_{4,o}$ lattice modes as well as $\eta_{\Gamma^{+}_{1,m}}$ and  $\eta_{\Gamma^{+}_{2,m}}$  macroscopic strains degrees of freedom.

The expansion coefficients have been adjusted on a training set of DFT data including 300 configurations (Fig.S6 and are reported in Table ST5). Amongst the various coupling terms, we find that the 3$^{rd}$ order coupling $Q_{\Gamma^{+}_{1,m}}Q_{\Gamma^{+}_{2,m}}^2$ is the most significant in lowering the energy (-456 meV/f.u.). Then, strain couplings $Q_{\Gamma^{+}_{2,m}} \eta_{\Gamma^{+}_{2,m}}$ (-100 meV/f.u.) and $Q_{\Gamma^{+}_{2,m}}^2 \eta_{\Gamma^{+}_{1,m}}$ (-232 meV/f.u.) are also significant. This highlights a rather complex and unusual phase transition mechanism in which many anharmonic couplings  of $\Gamma^{+}_{2,m}$ with $\Gamma^{+}_{1,m}$ ,  $\eta_{\Gamma^{+}_{1,m}}$ and $\eta_{\Gamma^{+}_{2,m}}$cooperate to lower the energy and produce the $P\bar{1}$ ground state.

\par
\begin{figure}[h!]
\includegraphics[width=\columnwidth]{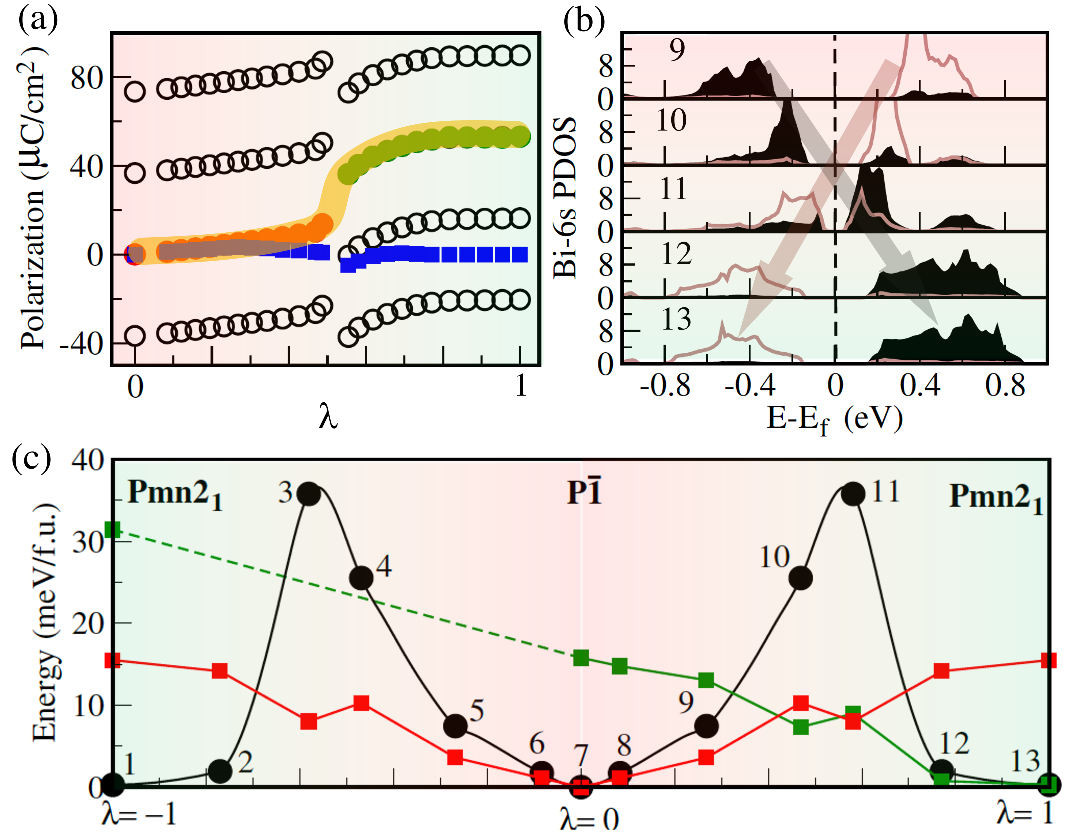}
\caption{(a) Evolution of $P^s_x$ along the insulating NEB path from $P\overline1$ ($\lambda=0$) to $Pmn2_1$ ($\lambda=1$), with CO evolving from C-type (red dots) to G-type (green dots). $P^s_y$ (blue dots) remains negligible along the path. (b)  Evolution of Bi 6s PDOS at selected points along the NEB path as defined in (c). Black and brown corresponds to Bi$_L$ and Bi$_s$ in $P\overline1$, which switches their character gradually. (c) Evolution of the energy along the NEB path connecting $P\overline1$ ($\lambda=0$) to $Pmn2_1$ with $+P^s_x$ ($\lambda=1$) and $-P^s_x$ ($\lambda=-1$): adiabatic path (black dots), non-adiabatic paths with atomic structure at  $\lambda=0$ (red dots) and $\lambda=+1$ (green dots) \cite{Nonadiabatic}.}
\label{CT}
\vspace{-0.8cm}
\end{figure}

{\bf Competing polar phase and electronic ferroelectricity} --
Being only 0.3 meV/f.u. higher in energy than the observed $P\bar{1}$ ground state, the $Pmn2_1$ phase emerges as a close and competing phase. As previously discussed, its G-type CO (Fig.\ref{CT}) phase breaks inversion symmetry, yielding a spontaneous polarization along $x$, $P_x^s$. Further, the direction of $P_x^s$ can be reversed by reversing the charge ordering (i.e. condensing  $\Gamma_{2,o}^-$ in opposite direction). Together, this makes $Pmn2_1$ a conceptual electronic ferroelectric phase, as long as experimental switching is practically achievable.
 
Estimating $P_x^s$ is not so trivial. Berry-phase calculation in the $Pmn2_1$ phase delivers a set of values  $P_x^s = -20.52 + n Q_P$ $\mu $C/cm$^2$  (with $n$ an integer and $Q_P= 36.76$ $\mu $C/cm$^2$ the polarization quantum), without clarifying which value of $n$ is appropriate. Using a Nudged Elastic Band (NEB) technique, we identified an insulating low-energy path  from non-polar $P\bar{1}$ to polar $Pmn2_1$ phase (with an energy barrier of 35 meV/f.u., Fig.\ref{metastable}a). From this, we can follow the evolution of $P_x^s$ along the path, as illustrated in Fig.\ref{CT}a. This shows first that the spontaneous polarization of the $Pmn2_1$ phase is $P_x^s = 53$ $\mu $C/cm$^2$, which is even larger than that of a conventional ferroelectric like BaTiO$_3$.  Then, it clarifies that the change of polarization is strongly non-linear with a jump of about 40 $\mu $C/cm$^2$. This jump that can be assigned to the change from C-type to G-type CO as highlighted from the PDOS of Bi in Fig.\ref{CT}b. It is also compatible (see SI) with the transfer of 2 electrons between Bi sites in one layer ($z=1/2$ in Fig.\ref{metastable}b) , confirming that $P^s_x$ mainly originates from electronic CO.   

The non-polar character of the $P\overline{1}$ ground state, combined with the very close energy of the $Pmn2_1$ ferroelectric phase ($\Delta E = 0.3$ meV/f.u.), makes BiNiO$_3$ a potential antiferroelectric. Applying an electric field ${\cal E}_{\rm{T}} = \Delta E / \Omega_0 P^s_x \approx 15$ kV/cm should be enough to stabilize thermodynamically $Pmn2_1$ against the $P\overline{1}$ phase. However, achieving electric field transition would a priori require a much larger field ${\cal E}_{\rm{A}}$ to overcome the adiabatic energy barrier between the two phases ($\Delta E_{\rm{ A}} \approx$ 35 meV/f.u. at zero field and zero kelvin). Alternatively, it might be questioned if non-adiabatic electron transfer would eventually be possible. Following the scheme proposed Qi and Rabe \cite{Karin_nonadiabatic_switching} (see Fig.\ref{CT}c and SI), we estimate the field required for non-adiabatic transition to ${\cal E}_{\rm{ NA}} = \Delta E_{\rm{ NA}} / \Omega_0 P^s_x \approx 800$ kV/cm ($\Delta E_{\rm{ NA}}=15$ meV/f.u.). As discussed by Qi and Rabe, this should not be taken as an exact value but rather as an estimate to compare distinct compounds. Our computed ${\cal E}_{\rm{ NA}}$ is larger than that in Fe$_3$O$_4$  \cite{Karin_nonadiabatic_switching} showing a similar bandgap and in which ferroelectric switching has been experimentally observed \cite{Fe3O4_Silvia}. It is however significantly smaller than in other electronic ferroelectrics like SrVO$_3$/LaVO$_3$ or LuFe$_2$O$_4$ \cite{Karin_nonadiabatic_switching}. As such, BiNiO$_3$ remains a plausible candidate for electronic antiferroelectricity, with field induced non-polar to polar transition potentially accessible and driven by non-adiabatic electron transfer.

{\bf Strain engineering} --  Interestingly, the lattice parameters (along a and c) of the $Pmn2_1$ phase are significantly different from those of the $P\overline{1}$ ground state (see Table.-ST7), which opens the perspective of using strain engineering to stabilize a ferroelectric ground state. It appears that the lattice parameters of the  NdGaO$_3$, a widely used substrate for the growth of perovskite oxide films, perfectly match with those of the $Pmn2_1$ phase. Comparing then the energies of different possible orientations of $P\overline{1}$ and $Pmn2_1$ phases, epitaxially strained on commercially available (110)$_o$ and (001)$_o$ NdGaO$_3$ substrates (Table.ST8,ST9), it appears that the ferroelectric $Pmn2_1$ phase is always elastically favored. In the case of the (110)$_o$ substrate, the strained $Pmn2_1$ ferroelectric phase lies 7 meV/f.u. below the strained $P\overline{1}$ and moreover aligns its long axis in plane as that of the substrate, which makes it a likely case to be realized experimentally.  Polarization switching in such a marginally strained polar $Pmn2_1$ phase would require reversing the CO and as such electron transfer in each of the two layers ($z = 0$ and $1/2$ of Fig.\ref{metastable}d). According to Fig.\ref{CT}c, it should be accessible from non-adiabatic electron transfer at the same reasonable field ${\cal E}_{NA} \approx  800$ kV/cm as before, making the system a potential electronic ferroelectric.            

{\bf Conclusions} --  
BiNiO$_3$  behaves differently  than other nickelate perovskites, which show CO at Ni sites. The charge transfer ${\rm Bi}^{3+}{\rm Ni}^{3+} \rightarrow {\rm Bi}^{4+}{\rm Ni}^{2+}$, yielding the ${\rm Bi}^{4+}$ valence skipper state, is the starting point for the electronic instability of the metallic $Pnma$ phase, which is then further stabilized by CO at Bi sites in the insulating non-polar $P{\bar 1}$ ground state and close feroelectric $Pmn2_1$ phase. 
TlMnO$_3$\cite{TlMnO} is another alternative perovskite we found  hosting a $P\overline{1}$ ground state. Interestingly, it also shows a metallic $Pnma$ to insulating $P\overline{1}$ phase transition but coming instead from orbital ordering at Mn$^{3+}$ sites. We want to stress that ferroelectricity in BiNiO$_3$ is distinct from that in other Bi$M$O$_3$ perovskites ($M$= Fe, Co, In) \cite{BiFeO3,BiCoO3,BiInO3} in which only Bi$^{3+}$ is present and polarisation driven by the lone pair of Bi$^{3+}$. In BiNiO$_3$, the polarisation arises from the G-type Bi$^{3+}$/Bi$^{5+}$ CO and is electronic in nature. Electronic ferroelectricty has been reported in non-perovskite Fe$_3$O$_4$\cite{Fe3O4_Silvia}, $A$Fe$_2$O$_4$ compounds \cite{LuFe2O4,YbFe2O4} and perovskite oxide superlattices \cite{KarinFe,KarinV} but remains a rare phenomena. Stabilizing the polar $Pmn2_1$ phase of BiNiO$_3$ by electric field or strain enginering appears as a promising new platform to probe further the intriguing concept of electronic (anti-)ferroelectricity.


$Acknowledgement:$ SB thanks He Xu for useful discussions and technical support. This work was supported  by F.R.S.-FNRS Belgium under PDR grant T.0107.20 (PROMOSPAN). The authors acknowledge the use of the CECI supercomputer facilities funded by the F.R.S-FNRS (Grant No. 2.5020.1) and of the Tier-1 supercomputer of the Fédération Wallonie-Bruxelles funded by the Walloon Region (Grant No. 1117545).


\begin{thebibliography}{42}
\bibitem [{Non()}]{Nonadiabatic}%
  \BibitemOpen
  \href@noop {} {\bibinfo {title} {Non-adiabatic paths have been obtained
  fixing the atomic geometry to $\lambda=0$ or $\lambda = 1$ and then
  constraining the electronic charge at each point to what it is along the
  adiabatic path, using constrained-$\mathrm{DFT}$ calculations.}}\BibitemShut
  {Stop}%
\end{thebibliography}

%
\end{document}